\documentclass[%
 aip,
 amsmath,amssymb,
preprint,%
]{revtex4-1}

\usepackage{graphicx}
\usepackage{dcolumn}
\usepackage{bm}

\usepackage[utf8]{inputenc}
\usepackage[T1]{fontenc}
\usepackage{mathptmx}

\begin{document}

\preprint{JCP}

\title[DCH-XES]{Double Core Hole Valence-to-Core X-ray Emission Spectroscopy: A Theoretical Exploration Using Time-Dependent Density Functional Theory}

\author{Yu Zhang}
 \affiliation{Stanford PULSE Institute, SLAC National Accelerator Laboratory, Menlo Park, California 94025, USA.}
 \email{yzhang15@slac.stanford.edu} 
 
 \author{Uwe Bergmann}
 \affiliation{LCLS and Stanford PULSE Institute, SLAC National Accelerator Laboratory, Menlo Park, California 94025, USA.}
 \email{bergmann@slac.stanford.edu}
 
  \author{Robert Schoenlein}
  \affiliation{SLAC National Accelerator Laboratory, Menlo Park, California 94025, USA.}
  
 \author{Munira Khalil}
 \affiliation{Department of Chemistry, University of Washington, Seattle, Washington 98195, USA}
 
 \author{Niranjan Govind}
 \affiliation{Environmental Molecular Sciences Laboratory, Pacific Northwest National Laboratory, Richland, WA 99352, USA.}
 \email{niri.govind@pnnl.gov}
 

\date{\today}

\begin{abstract}
With the help of newly developed X-ray free-electron laser (XFEL) sources, creating double core holes simultaneously at the same or different atomic sites in a molecule has now become possible. Double core hole (DCH) X-ray emission  is a new form of X-ray nonlinear spectroscopy that can be studied with a XFEL. Here we computationally explore the metal K-edge valence-to-core (VtC) X-ray emission spectroscopy (XES) of metal/metal and metal/ligand double core hole states in a series of transition metal complexes with time-dependent density functional theory. The simulated DCH VtC-XES signals are compared with conventional single core hole (SCH) XES signals. The energy shifts and intensity changes of the DCH emission lines with respect to the corresponding SCH-XES features are fingerprints of the coupling between the second core hole and the occupied orbitals around the DCHs that contain important chemical bonding information of the complex. The core hole localization effect on DCH VtC-XES is also briefly discussed. We theoretically demonstrate that
DCH XES provides subtle information on the local electronic structure around metal centers in transition metal complexes beyond conventional linear XES. Our predicted changes from calculations between SCH-XES and DCH-XES features should be detectable with modern XFEL sources.
\end{abstract}

\maketitle


\section{Introduction}
One of the striking effects of intense X-ray-matter interaction is the creation of multiple core holes. Theoretical insight into this phenomenon was provided long before any realistic experiments on DCH states were performed. More than three decades ago, Cederbaum et al. studied DCHs in small molecules theoretically, and predicted that the electron binding energies associated with DCHs at different atomic sites could sensitively probe the chemical environment of the ionized atoms\cite{Cederbaum86}. Since then, there have been several theoretical DCH studies and related spectroscopic signals with various methods including many-body Green's function\cite{Ohrendorf91,Santra09,Kryzhevoi11}, multiple configuration self-consistent field (MCSCF)\cite{Agren93,Tashiro10,Tashiro10b,Takahashi11,Takahashi11b,Tashiro12,Takahashi15,Hua16,Koulentianos18}, density functional theory (DFT)\cite{Takahashi11,Takahashi11b,Takahashi11c,Ueda12,Thomas12,Takahashi14,Takahashi15,Takahashi18}, time-dependent density functional theory (TDDFT)\cite{Zhang13}, Møller–Plesset perturbation theory\cite{Thomas12}, and the $Z+1$ approximation\cite{Schweigert07}, respectively.

DCH spectroscopy was originally studied with synchrotron 
radiation,\cite{Linusson11,Lablanquie11,Lablanquie11b,Penent12,Nakano13,Carniato15,Penent15} where the absorption of one photon is accompanied by the ejection of two core electrons. In this process the correlation between the two ejected core electrons plays an important role. While multi-photon processes with synchrotron radiation is unlikely, it becomes possible with more intense X-ray free-electron laser (XFEL) pulses. Thanks to the rapid development of XFEL sources, two-photon photoelectron spectroscopy or double core hole (DCH) spectroscopy has been shown to be a powerful tool to probe the chemical environment of specific atomic sites in molecules\cite{Berrah11,Salen12, Ueda12, Piancastelli13, Takahashi15, Berrah15}. 
Recently, DCH states have been created using XFEL pulses at LCLS in neon\cite{Young10}, nitrogen gas\cite{Cryan10, Fang10, Salen12}, N$_2$O, CO$_2$ and CO\cite{Salen12}, and the aminophenol molecule\cite{Zhaunerchyk15}. Differences between the photoelectron spectral data taken with focused and unfocused laser beams were studied in order to extract the DCH contribution to the signal\cite{Salen12}. It is believed that two-site DCH (ts-DCH) spectroscopy provides a more sensitive probe for the local chemical environment of the excited atoms than does single core hole spectroscopy\cite{Salen12,Piancastelli13}. Mukamel et al. theoretically studied X-ray four-wave mixing spectroscopy involving DCHs\cite{Schweigert07,Zhang13,Hua16}. The corresponding experiments require well-controlled intense X-ray pulses which are not currently available. In this study, we focus on ts-DCH spectroscopy because it is more sensitive for chemical analysis than single-site DCH spectroscopy. Previous DCH spectroscopy experiments measured photoelectrons\cite{Berrah11,Salen12,Larsson13,Mucke15,Koulentianos18b} and Auger electrons\cite{Eland10,Cryan10,Fang10,Penent12,Mucke15,Goldsztejn16}, which are not suitable for solution samples common in chemistry. We propose the use of intense XFEL pulses to create DCHs at different sites in the system, and to study the corresponding X-ray emissions. Compared with conventional X-ray emissions with single core holes (SCHs), X-ray emissions with DCHs probe the occupied orbitals perturbed by the additional core hole in the system, and thus carry information about the local electronic structure as well as the coupling between the two core holes. 


In this paper, we study ts-DCH spectroscopy from a theoretical standpoint using TDDFT. We calculate the VtC-XES\cite{Gallo14,Bauer14} signals resulting from DCHs of a series of mono- and binuclear transition metal model complexes.  
VtC-XES signals carry more chemical information of the system than K$\alpha$ and K$\beta$ mainline emissions, because they directly probe the valence orbitals. For the binuclear complexes with metal-metal direct bonds, VtC-XES signals of the metal 1s DCHs at different sites were studied. Mononuclear complexes with different Mn oxidation states (II, III) have been used to investigate the emissions of the metal-1s/ligand-1s DCH states, from which the information on the chemical bonds between the metal center and the coordinating atoms are revealed.  


\section{Computational Details}
\label{sec:compt}
All calculations were performed at the DFT and TDDFT (within the Tamm-Dancoff approximation
(TDA)\cite{Hirata99}) levels of theory with the NWChem quantum chemistry package.\cite{NWChem} No symmetry restrictions have been applied. The PBE0\cite{Perdew96,Adamo99} hybrid functional was used for all calculations. All geometries were optimized at the PBE0/Def2-TZVP\cite{Weigend05} level of theory. The X-ray emission calculations have been performed using the FCH(full core hole)/TDDFT approach described in our previous publication.\cite{Zhang15} Here, this approach has been extended to explore double core hole states. In our SCH and DCH signal simulations, both $\alpha$ and $\beta$ ionization channels and all of their possible combinations were considered with equal weights. As an illustration a sample NWChem input file with notes is provided in Supplementary Material.  For the binuclear complexes, all calculations have been performed in the gas phase, while for the Mn mononuclear complexes, all calculations were performed in the CH$_2$Cl$_2$ solution phase ($\epsilon = 9.08$), which is described by the conductor-like screening model (COSMO)\cite{Klamt93}. For the TDDFT calculations of the core hole states, the Sapporo-TZP-2012\cite{Noro12}/Def2-TZVP basis sets have been used for the metal and non-metal atoms, respectively. In the localized N1s core hole calculations, in order to fix a 1s core hole at an individual N atom, we use an all-electron basis set representation only on one N atom and use effective core potentials (ECP) and the corresponding basis sets to describe the other N atoms. Specifically in this study, the Def2-TZVP basis set has been used for the N atom with a 1s core hole, and the SBKJC ECP\cite{Stevens84} together with the corresponding polarized basis set\cite{Labello06} has been used for the other N atoms without 1s core holes. In the SCH and DCH calculations of the studied binuclear complexes, transitions to core holes at both metal atoms are included.

\section{Results and discussion}
 \subsection{Binuclear transition metal complexes}
We have studied the VtC-XES signals of metal 1s/metal 1s ts-DCH states of binuclear transition metal complexes with strong metal-metal bonds. Polynuclear transition metal complexes with direct metal-metal bonds\cite{CottonBook} have attracted the attention of chemists for a long time, not only for their high bond orders (>3), but also for their important applications in making metal-organic frameworks, molecular conductors, photosensitizers and catalysts\cite{Berry17}. Results from theoretical calculations on binuclear metal complexes with metal-metal bonds have been reviewed recently \cite{Lyngdoh18}. Here, we chose one Fe complex Cp$_2$Fe($\mu$-CO)$_2$ (Fe-cplx, Cp = cyclopentadienyl) and one Co complex (C$_4$H$_6$)$_2$Co$_2$($\mu$-CO) (Co-cplx) as candidate systems. The structures of the two complexes can be found in Fig. \ref{fig:Fe2} and \ref{fig:Co2}. The optimized Fe\---Fe and Co\---Co bond lengths are 2.130 and 2.110 \AA{}, respectively. These bond lengths are close to previously reported theoretical values (2.120 \AA{} for Fe\---Fe and 2.142 \AA{} for Co\---Co) \cite{Wang06,Fan12}. 
Bond valence analysis shows that Fe-cplx has a formal bond order of 3\cite{Wang06} and Co-cplx has a formal bond order of 4\cite{Fan12}, respectively.

In general, compared with SCH states, the second core hole in ts-DCH states have two types of effects on XES: 1) emission energy shifts due to the additional attractive potential of the second core hole; 2) emission intensity changes due to additional perturbations of the molecular orbitals caused by the second core hole. An electron in a localized orbital near the second deep 1s core hole will feel a very strong attraction, resulting in a blue shift of the corresponding emission energies. For other orbitals not localized near the second core hole, the attraction and screening of the core hole can have an effect on the overall orbital shape. On the other hand, if the two core holes are uncorrelated or independent, both emission energy and intensity changes will be negligible and should result in a spectrum similar to the corresponding SCH-XES spectrum.
Comparing the emission energies and intensities (Figs. \ref{fig:Fe2} and \ref{fig:Co2}) allows one to shed light on the strength of the interaction between the two core holes. 

\subsubsection{Fe-cplx}
\begin{figure}
 	\centering
 	\includegraphics[width=0.7\linewidth]{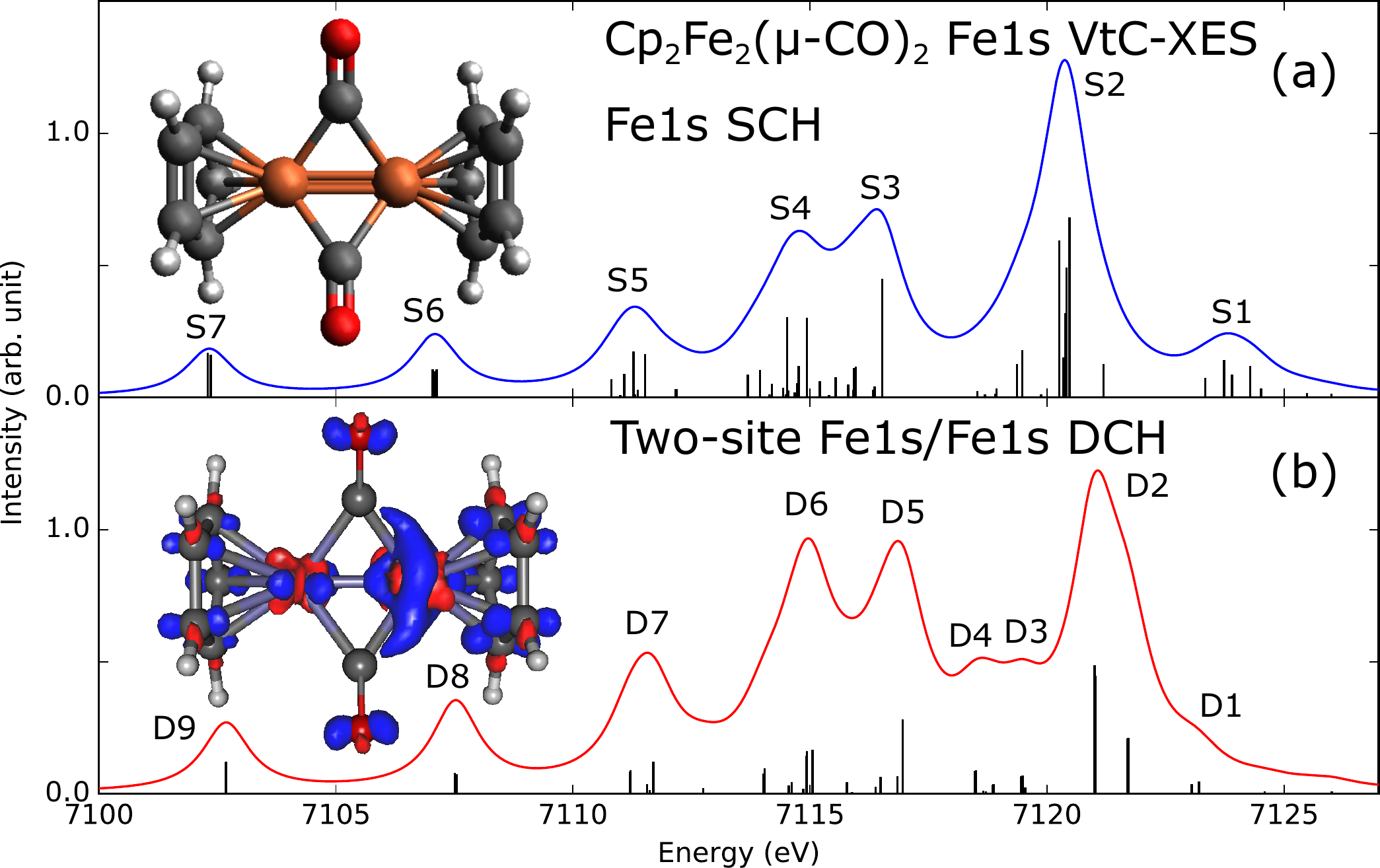}
 	\caption[Calculated SCH and DCH Fe1s VtC-XES signals of Cp$_2$Fe($\mu$-CO)$_2$ (Fe-cplx, Cp = cyclopentadienyl).]{Calculated SCH and DCH Fe1s VtC-XES signals of Cp$_2$Fe($\mu$-CO)$_2$ (Fe-cplx, Cp = cyclopentadienyl). The calculated spectra have been Lorentzian broadened by 1.2 eV. SCH and DCH spectra are scaled differently for the convenience of plotting. Stick heights in different panels are not calibrated. (a) Calculated SCH Fe1s VtC-XES signals. Important features are labeled as S1\---7. The molecular structure is also shown. Color code: brown, Fe; red, O; dark grey, C; light grey, H. (b) Calculated two-site Fe1s/Fe1s DCH Fe1s VtC-XES signals. Important features are labeled as D1\---9. The calculated electron density difference between the DCH and the SCH state is also shown ($\rho_{\textbf{DCH}}-\rho_{\textbf{SCH}}$, surface isovalue = 0.005). Red and blue denotes positive and negative values, respectively. }
 	\label{fig:Fe2}
 \end{figure}

In Fig. \ref{fig:Fe2} (b), we show the calculated electron density difference between the DCH and the SCH states of Fe-cplx ($\rho_{\textbf{DCH}}-\rho_{\textbf{SCH}}$). One can clearly see that the hole on the right Fe atom induces significant electron density redistribution (blue means hole and red means particle density). There are also some p-type electron density changes on the Cp ring and O atom of the CO ligand, which suggests that emission involving orbitals with similar character could be significantly affected by the second core hole. For  Fe-cplx, both the SCH and DCH spectra have a shoulder feature above 7123 eV (labeled as S1 in Fig. \ref{fig:Fe2}(a) and and D1 in panel (b)), but the DCH-XES peak is red-shifted by $\sim 0.5$ eV and much weaker (relative to the strongest peak) compared with the corresponding SCH-XES feature. 

Molecular orbital (MO) analysis of the representative transitions of the two peaks shows that the largest contribution of each comes from the Fe\---Fe d $\pi$ orbitals (see Table S1 in Supplementary Material for the plots of the MOs discussed for Fe-cplx). However, for the SCH-XES transition, the dominant MOs are at the Fe atoms, while for the DCH-XES transition, MOs on the Cp ring are also involved because of the second core hole. The involvement of the Cp ring orbitals may explain the reduced intensity of the corresponding transition in the DCH-XES spectrum. For the strongest SCH-XES features around 7120.4 eV (S2 and D2 in Fig. \ref{fig:Fe2}), the DCH-XES peak D2 is blue-shifted by $\sim 0.7$ eV compared to S2. Local Fe d orbitals and Fe\---Fe d $\pi$ orbitals contribute significantly to such transitions, and the influence of the second core hole on these transitions is manifest in the energy shift.

The DCH-XES spectrum has relatively stronger shoulder features between 7118 and 7120 eV (D3 and D4 on the high platform in Fig. \ref{fig:Fe2}(b)), respectively. These transitions involve local Fe d orbitals and Fe\---Fe d $\sigma$ bonding orbitals, which are also heavily affected by the second core hole. These emission lines contain information about the Fe\---Fe direct bonding, but cannot be clearly resolved in the SCH-XES spectrum as it is suppressed by the strongest peak S2 nearby. The S3 peak mainly represents transitions from orbitals on the CO ligands, while the D5 peak has many transition components from the Cp ring C p orbitals, which again is the effect of the second core hole. Compared to S3, D3 is blue-shifted by $\sim 0.5$ eV. All the other peaks (S4-7 and D6-9) have dominant transition contributions from the orbitals on the CO and Cp ligands, which are not heavily affected by the second core hole, and thus similar in both the SCH-XES and DCH-XES spectra.
From the analysis above, we see that the second Fe1s core hole energy shifts and intensity changes in the higher energy K$\beta_{2,5}$ region are affected to a greater extent compared with the lower energy K$\beta"$ features. 
This is probably because that the K$\beta_{2,5}$ features involve MOs with more metal d orbital character than those involved by the K$\beta"$ features.


\subsubsection{Co-cplx}
\begin{figure}
 	\centering
 	\includegraphics[width=0.7\linewidth]{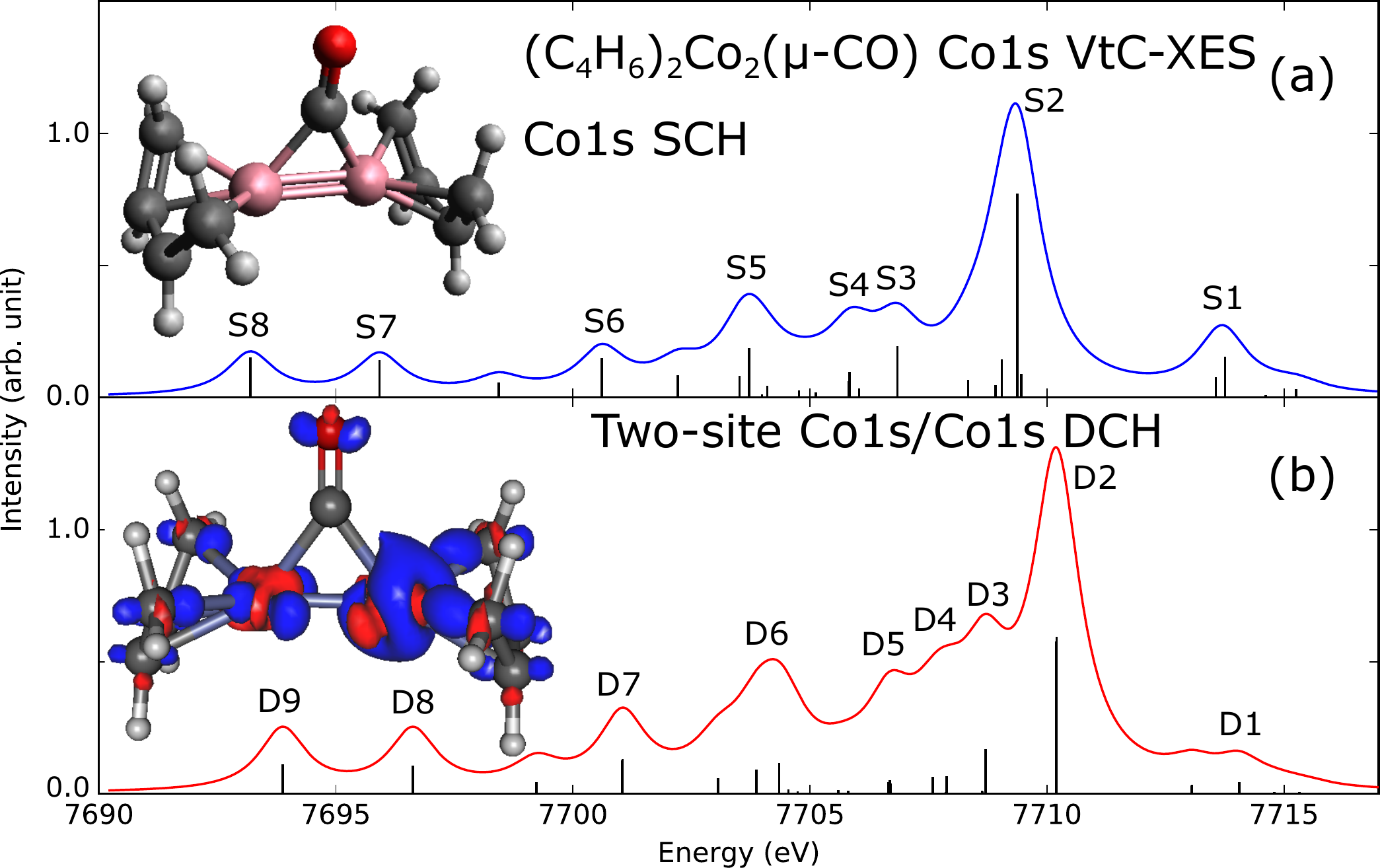}
 	\caption[Calculated SCH and DCH Fe1s VtC-XES signals of (C$_4$H$_6$)$_2$Co$_2$($\mu$-CO) (Co-cplx).]{Calculated SCH and DCH Fe1s VtC-XES signals of (C$_4$H$_6$)$_2$Co$_2$($\mu$-CO) (Co-cplx). All calculated spectra have not been shifted, and been Lorentzian broadened by 1.2 eV. SCH and DCH spectra are scaled differently for the convenience of plotting. Stick heights in different panels are not calibrated. (a) Calculated SCH Co1s VtC-XES signals. Important features are labeled as S1\---8. The molecular structure is also shown. Color code: pink, Co; red, O; dark grey, C; light grey, H. (b) Calculated two-site Co1s/Co1s DCH Co1s VtC-XES signals. Important features are labeled as D1\---9. The calculated electron density difference between the DCH and the SCH state is also shown ($\rho_{\textbf{DCH}}-\rho_{\textbf{SCH}}$, surface isovalue = 0.006). Red and blue denotes positive and negative values, respectively. }
 	\label{fig:Co2}
 \end{figure}

In Fig. \ref{fig:Co2} (b), we also show the calculated electron density difference between the DCH and the SCH state of Co-cplx ($\rho_{\textbf{DCH}}-\rho_{\textbf{SCH}}$). Similar to the case of Fe-cplx, one can see the hole on the right Co atom induces electron density changes around it and some p-type electron density deficiency on the C$_4$H$_6$ ligand and O atom of the CO ligand. Peak S1 is much stronger (relative to the strongest peak) than D1 (see Fig. \ref{fig:Co2} for labeling). An inspection of the contributing MOs involved in the transitions (see Table S2 in Supplementary Material) suggests that S1 involves localized Co d orbitals bonded with p orbitals on the C$_4$H$_6$ ligand only around the SCH, while peak D1 involves localized Co d orbitals of both Co atoms, which reduces its emission intensity. The strongest DCH-XES peak D2 is blue-shifted by about 0.8 eV compared to the strongest SCH-XES peak S2. This shift is even larger than that of Fe-coplx ($\sim 0.6$ eV) and should be detectable at the current levels of instrumentation.  Both transitions involve d orbitals at two Co sites interacting (bonding or anti-bonding) with the C p orbitals on the C$_4$H$_6$ ligand. D3-5 form a rising shoulder beside the strongest peak D2. Similar to the case of Fe-cplx, D4 contains the information of direct Co\---Co bonding and the corresponding peak cannot be resolved in the SCH-XES spectrum. D6 is blue-shifted by around 0.5 eV compared to S5. Their transitions have both significant contributions from CO p $\pi$ orbitals and $\sigma$ bonding orbitals on the C$_4$H$_6$ ligands. Both S6 and D7 mainly represent emissions from C and O 2s orbitals on the CO ligand. D7 is blue-shifted by about 0.4 eV compared to S6. Going further to the lower energy range, D8 is shifted by $\sim 0.7$ eV compared to S7 and D9 is shifted by $\sim 0.6$ eV compared to S8. All of these can be considered as transitions from the C 2s orbitals from the C$_4$H$_6$ ligands. Unlike  the Fe-cplx, for the Co-cplx, it seems that both K$\beta_{2,5}$ and K$\beta"$ emission lines are shifted in the DCH-XES spectrum compared to the corresponding SCH-XES. As K$\beta"$ emissions mainly come from ligand orbitals, the shifting of these lines tells us in Co-cplx the metal-ligand interaction is stronger than that in Fe-cplx, and DCH-XES may contain coordination chemical information beyond SCH-XES.

 \subsection{Mn mononuclear complexes}
 \label{sec:Mn-mono}
 
In the last section we studied the case of the VtC-XES of two-site metal 1s/metal 1s DCH states, which in principle can be created by a single intense hard X-ray pulse with enough brilliance. Here, we propose creating metal 1s/ligand 1s DCH states in transition metal complexes, which may be achieved by combining hard and soft X-ray pulses. Using typical Mn mononuclear complexes as candidate systems, we investigate the VtC-XES of Mn1s/ligand 1s DCH states. Compared with metal core holes, core holes on ligands might have greater impact on the valence orbitals around the metal center and therefore the corresponding VtC-XES spectra could be more informative about the coordination bonds.
 
\begin{figure}
 	\centering
 	\includegraphics[width=0.7\linewidth]{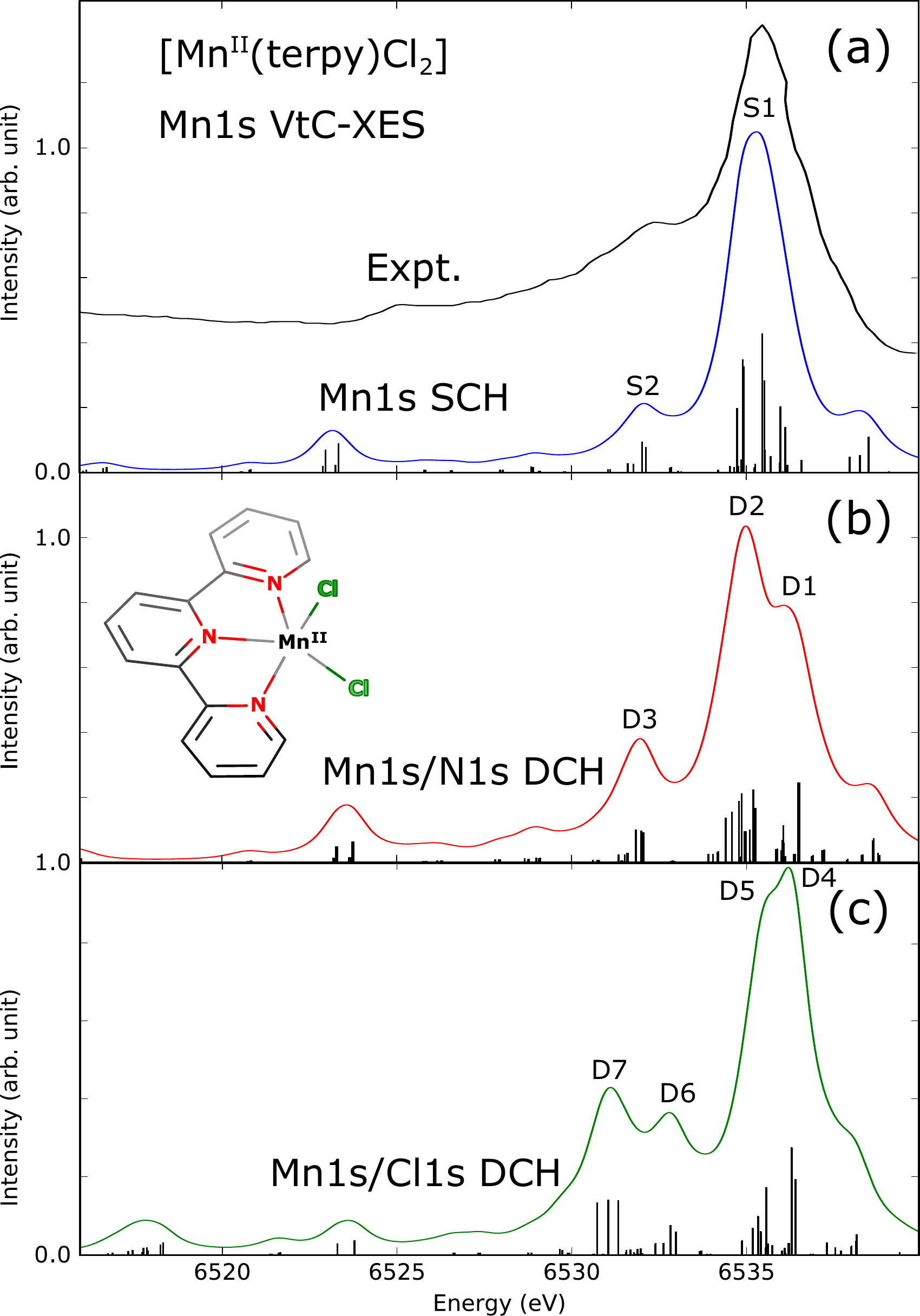}
 	\caption[Calculated SCH and DCH Mn1s VtC-XES signals of Mn(II)(terpy)Cl$_2$.]{Calculated SCH and DCH Mn1s VtC-XES signals of [Mn(II)(terpy)Cl$_2$] (Mn\_II-cplx). All calculated spectra have been red-shifted by 14.07 eV to be compared with the conventional SCH VtC-XES experiment\cite{Beckwith13}, and been Lorentzian broadened by 1.2 eV. SCH and DCH spectra are scaled differently for the convenience of plotting. Stick heights in different panels are not calibrated. (a) Experimental and calculated SCH Mn1s VtC-XES signals. Important features are labeled as S1, S2. (b) Calculated Mn1s/N1s DCH Mn1s VtC-XES signals. Important features are labeled as D1\---3. The molecular structure of Mn\_II-cplx is also shown. (c) Calculated Mn1s/Cl1s DCH Mn1s VtC-XES signals. Important features are labeled as D4\---7. }
 	\label{fig:Mn_II}
 \end{figure}

The calculated Mn1s SCH and Mn1s/N1s and Mn1s/Cl1s DCH VtC-XES signals of a representative high-spin Mn(II) mononuclear  complex [Mn(II)(terpy)Cl$_2$] (Mn\_II-cplx, terpy = 2,2';6',2"-terpyridine, sextet) at Mn K-edge are presented in Fig. \ref{fig:Mn_II}. We note that in all the Mn1s/Cl1s DCH calculations, the Cl1s core hole is localized on one of the Cl atoms, while for the  Mn1s/N1s DCH calculations, the N1s core hole is delocalized to all N atoms. The issue of localized/delocalized core hole in DCH calculations will be addressed in the next section. We focus on the important features labeled in the figure. S1 around 6535.4 eV is the strongest peak in the conventional SCH VtC-XES spectrum, of which the major contributing occupied MO has significant components as the Mn\---N coordination bond and sigma bonds on the pyridine rings (see Table S3 in Supplementary Material). S2 at ~ 6532.0 eV also represents transitions from orbitals on the pyridine rings. For the Mn1s/N1s DCH spectrum, the shoulder peak D1 around 6536.0\---6535.5 eV denotes transitions from the Mn\---Cl bonds and Cl 3p orbitals. This peak cannot be resolved in the SCH spectrum. It is interesting that the N1s core hole favors the transition from the Cl atoms.
 The strongest peak D2 in panel (b) has the same character as S1 (see Table S3), but is red-shifted by $\sim$  0.5 eV. This is the effect of the N1s core hole. However, the N1s core hole  has almost no effect on D3 since it is essentially the same as S2. The Mn1s/Cl1s spectrum (panel (c) in Fig. \ref{fig:Mn_II}) is quite different from the other two spectra in the figure. The strongest peak D4 is of the same character as S1, but is blue-shifted by about 0.8 eV. The small shoulder D5 is in the same energy range of S1 but of totally different character: it mainly represents transitions from the Mn\---Cl coordination bonding orbitals (the Cl atom has not a core hole).  D6 at $\sim$ 6532.8 eV is similar to S2 in character, but is blue-shifted by $\sim$ 0.8 eV. The relatively strong peak D7 at $\sim$ 6531.1 eV represents transitions from 3p orbitals of the Cl with the 1s core hole, which is very different from D3 and S2 in character.

\begin{figure}
 	\centering
 	\includegraphics[width=0.7\linewidth]{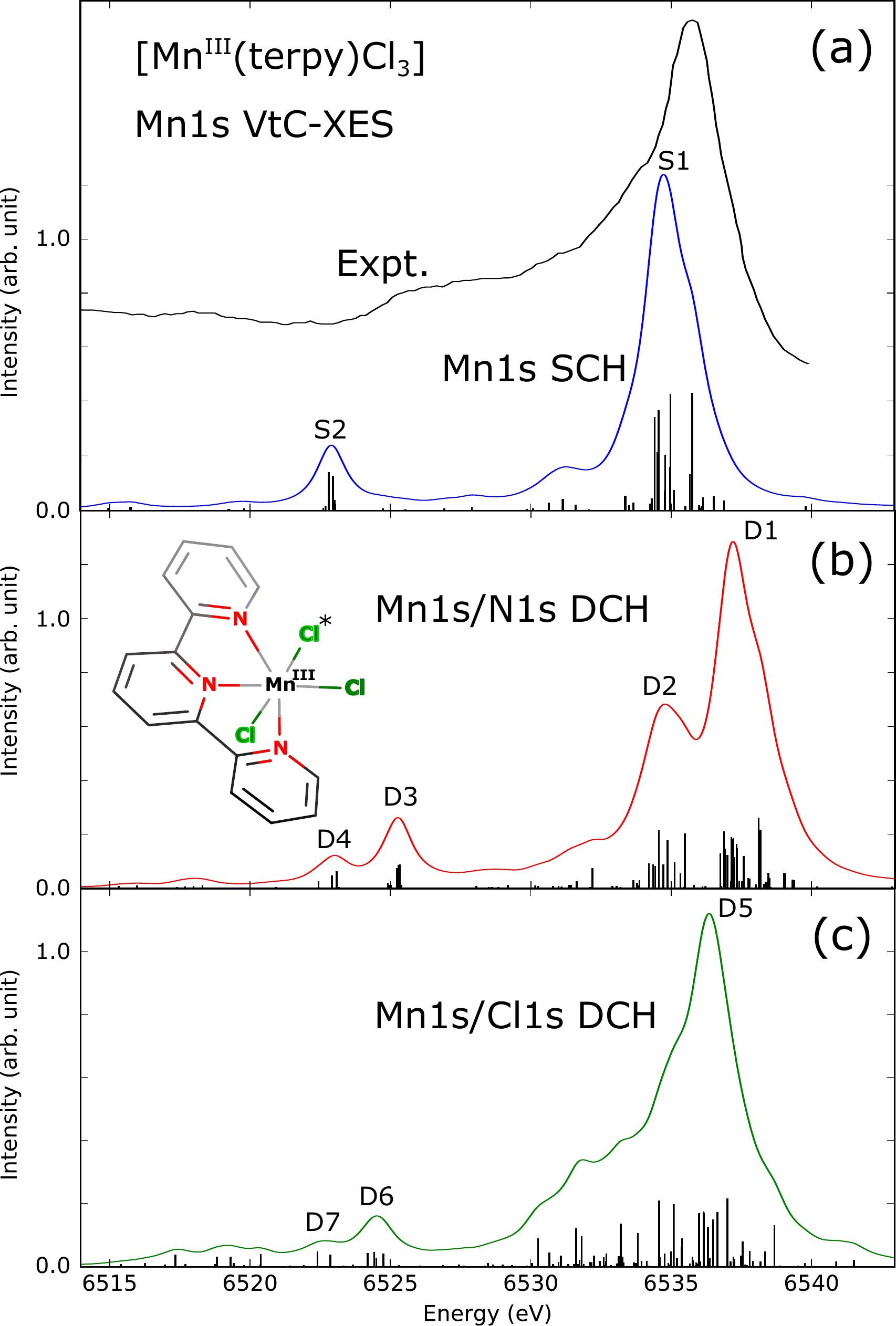}
 	\caption{Calculated SCH and DCH Mn1s VtC-XES signals of [Mn(III)(terpy)Cl$_3$](Mn\_III-cplx). All calculated spectra have been red-shifted by 12.04 eV to be compared with the conventional SCH VtC-XES experiment\cite{Beckwith13}, and been Lorentzian broadened by 1.2 eV. SCH and DCH spectra are scaled differently for the convenience of plotting. Stick heights in different panels are not calibrated. (a) Experimental and calculated SCH Mn1s VtC-XES signals. Important features are labeled as S1, S2. (b) Calculated Mn1s/N1s DCH Mn1s VtC-XES signals. Important features are labeled as D1\---4. The molecular structure of Mn\_III-cplx is also shown. The asterisk symbol on the Cl atom denotes Cl1s core hole site. (c) Calculated Mn1s/Cl1s DCH Mn1s VtC-XES signals. Important features are labeled as D5\---7.}
 	\label{fig:Mn_III}
 \end{figure}

For comparison, we study another high-spin Mn mononuclear complex [Mn(III)(terpy)Cl$_3$] (Mn\_III-cplx, quintet). The calculated Mn1s SCH and Mn1s/N1s, Mn1s/Cl1s DCH VtC-XES signals at Mn K-edge are shown in Fig. \ref{fig:Mn_III}.  As indicated in panels (a) and (b), the Mn1s/N1s DCH VtC-XES signals are very different from the Mn1s SCH counterpart. The strongest peak D1 is blue-shifted by about 2.5 eV compared to S1! A MO analysis reveals that the strong transitions in the broad peak S1 mainly involve MOs on the two side pyridine rings, while strong transitions in D1 and D2 have significant contributions from the Cl 3p orbitals and the MOs on the middle pyridine ring (see Table S4 in Supplementary Material). This D1 peak resembles the D1 peak in Fig. \ref{fig:Mn_II} in MO character. The huge shift of D1 indicates that the N1s core hole drastically changes the local electronic structure around the Mn metal center, and there is much stronger strong hybridization between the N 2p and Mn 3d orbitals in Mn\_III-cplx compared to Mn\_III-cplx. Peak D4 is similar to S2, both represent transitions from Cl 3s orbitals. D3 features with energies more than 2 eV higher than that of S2 also represent transitions from Cl 3s orbitals, but it is not seen in the SCH spectrum and should be considered as the effect of the N1s core hole. Because there are two chemically non-equivalent Cl atoms in this complex, we chose to put the 1s core hole at one of the Cl atoms perpendicular to the terpy plane, as labeled with an asterisk symbol in panel (b) of  Fig. \ref{fig:Mn_III}. The Mn1s/Cl1s DCH spectrum has a very broad shoulder on the lower energy side of the strongest peak D5, lacking characteristic features. The strong transitions above 6534 eV mainly involve Cl 3p orbitals, Mn\---Cl and Mn\---N coordination bonding orbitals. The weak peak D7 is similar to D4 and  S2. D6 resembles D3 in MO character but is red-shifted by $\sim$ 0.8 eV because its transition orbital is on the Cl atom with a core hole. 

From this analysis, we see that metal-ligand DCH VtC-XES can have shifted or new features compared to conventional SCH VtC-XES. These shifted or new features potentially contain additional chemical information of the coordination bonds between the metal center and ligands. 

 \subsection{Localized and delocalized core hole models}
 When studying core holes on multiple chemically equivalent atoms in a molecule a question arises whether the core hole should be considered as localized on one atom or delocalized to all equivalent atoms. Chemically equivalent atoms are symmetric and a localized core hole on one of them breaks the symmetry, leading to the L\"{o}wdin's symmetry lemma of Hartree-Fock theory\cite{Lowdin63}. This core hole localization and symmetry breaking issue has been raised and discussed in theoretical and experimental studies for decades. Bagus and Schaeffer discovered that in diatomic molecules the error of core ionization potentials from Hartree-Fock calculations could be greatly reduced if a localized core hole model, rather than a delocalized core hole model, is used\cite{Bagus72}. Cederbaum and Domcke pointed out that in a decomposition of the core ionization potential, the relaxation energy contribution is much larger than the correlation energy contribution if a localized core hole model is used, while in a delocalized core hole model, relaxation and correlation effects are both important\cite{Cederbaum77}. This can be used to explain why the localized core hole model works very well for calculating core ionization potentials with independent particle theories such as Hartree-Fock. Since these two seminal studies, more evidence favoring localized models in determining core hole properties has emerged\cite{Dill78,Muller79,Agren81b,Arneberg82,Chong07}. The symmetry breaking issue can be remedied by employing high-level electron correlation methods such as MCSCF methods\cite{Bacskay87}. A specific double excitation configuration describing a core-core excitation coupled to a valence-valence excitation under the symmetry restriction helps to reduce the symmetry breaking relaxation error\cite{Carravetta13}. In experiment both localized and delocalized core holes can be selectively detected\cite{Schoffler08,Guillemin15}.
 
A complete comparison of localized/delocalized core hole models in DCH spectroscopy goes beyond the scope of this paper. Here, we  present only a special case of localized/delocalized ligand core hole models in metal 1s/ligand 1s DCH VtC-XES. During the course of our calculations reported in Section \ref{sec:Mn-mono}, we found that for the 1s core holes on light atoms such as N, self-consistent field (SCF)  calculations often converge to a delocalized core hole state in which the core hole is almost equally distributed over all N atoms in the molecular  complex. We note that in our complexes the N atoms are not chemically equivalent and this is an example of  hole localization without strict symmetry. However, for deeper Cl1s and metal 1s core holes, we did not see core hole delocalization in the SCF calculations without symmetry constraints. In order to steer the SCF calculation to our target localized core hole state, we must freeze all the N1s electrons but one. Our strategy is to apply pseudopotentials to represent all N 1s electrons but the target one. See Section \ref{sec:compt} for computational details.
 
\begin{figure}
 	\centering
 	\includegraphics[width=1.0\linewidth]{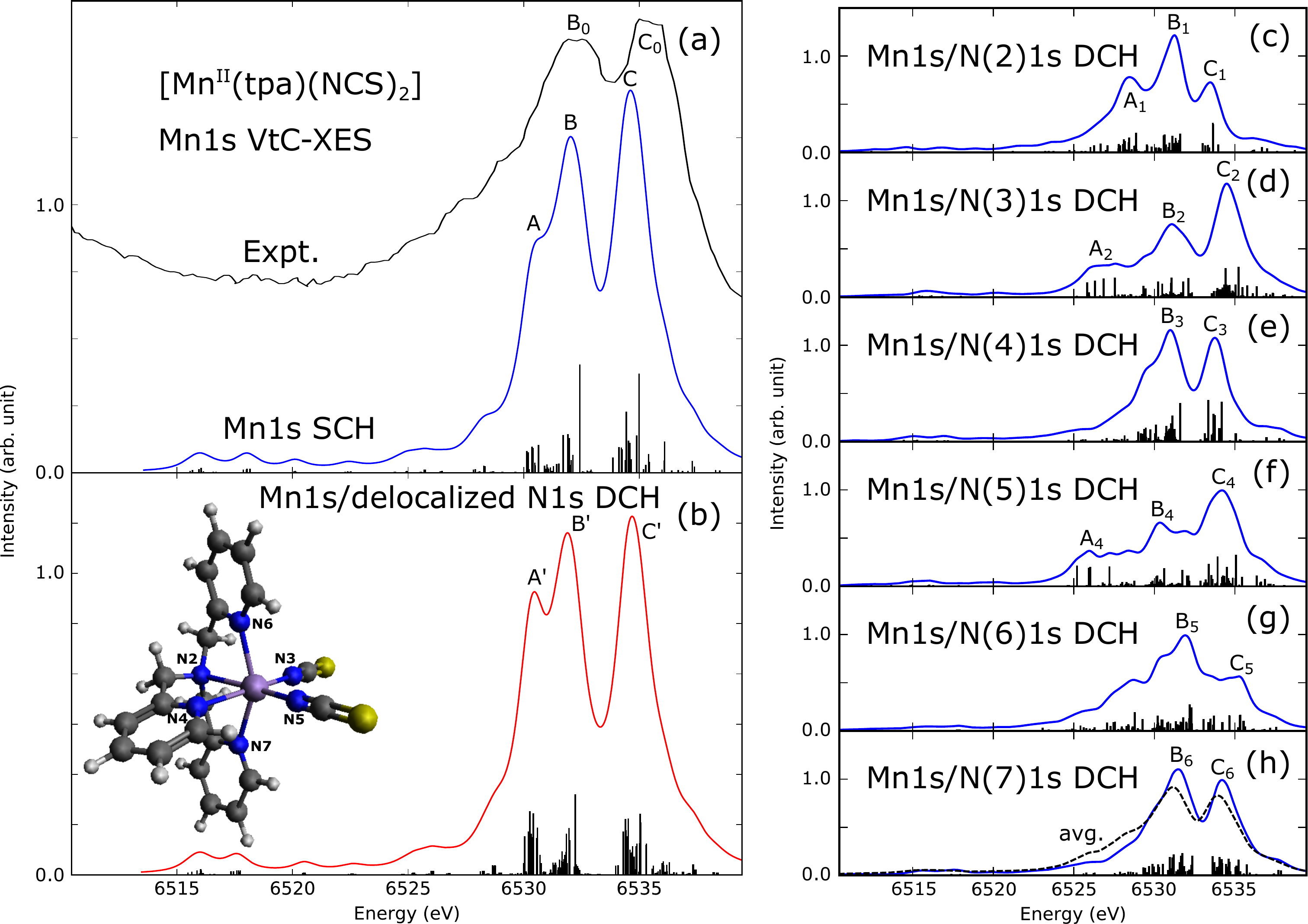}
 	\caption{Calculated Mn1s/N1s DCH Mn1s VtC-XES signals of [Mn(II)(tpa)(NCS)$_2$] (Mn\_tpa-cplx, tpa = Tris(2-pyridylmethyl)amine) from delocalized and localized core hole models. All calculated spectra have been red-shifted by 14.55 eV to be compared with experiment\cite{Beckwith13}, and been Lorentzian broadened by 1.2 eV. SCH and DCH spectra are scaled differently for the convenience of plotting. Stick heights in different panels are not calibrated. (a) Experimental and calculated SCH Mn1s VtC-XES signals. (b) Calculated Mn1s/N1s DCH Mn1s VtC-XES signals with N1s core hole delolalized to all N atoms. The molecular structure and N atom labeling is also shown. Color code: purple, Mn; blue, N; yellow, S; dark grey, C; light grey, H. (c)\---(h) Calculated Mn1s/N1s DCH Mn1s VtC-XES signals with the N1s core hole localized at the specific N atoms as labeled on the molecular structure in panel (b). The dotted curve in panel(h) represent the average of all the spectra in panel (c)\---(h).  }
 	\label{fig:Mn_II-tpa}
 \end{figure}

The calculated Mn1s/N1s DCH Mn1s VtC-XES signals of [Mn(II)(tpa)(NCS)$_2$] (Mn\_tpa-cplx) using the delocalized and localized core hole models are shown in Fig. \ref{fig:Mn_II-tpa}. We chose this complex for our study because it has only N atoms in its coordination sphere. We note that we did not impose any symmetry in our calculations, so all the N atoms (including N6 and N7 in panel(b)) are not equivalent in our optimized geometry. As we described above, the signals in Fig. \ref{fig:Mn_II-tpa} panels (a) (b) were calculated with all-electron basis sets and the signals in panels (c)\---(h) were calculated with pseudopotential basis sets, a direct quantitative comparison on the signals from the delocalized/localized core hole models may be misleading. Therefore we focus on the spectral profiles in this section. From Fig. \ref{fig:Mn_II-tpa} (a) and (b) we can see despite the intensity difference between the shoulder peaks A and A', the Mn1s/delocalized N1s DCH VtC-XES signal is very similar to the conventional Mn1s SCH VtC-XES signal. However, the spectral profiles of Mn1s/localized N1s DCH VtC-XES signals (panels (c)\---(h) are different from those in panels (a) and (b). This is understandable since an 1s core hole localized to an individual N atoms would induce a very different electron density redistribution from that caused by a more spherical delocalized N1s core hole. We also notice that in the localized core hole model different N1s core holes lead to different DCH VtC-XES signals, which may be used to probe the physical occurrence of localized DCH ionizations. An easy inspection of the curves in panels (c)\---(h) groups the N atoms into 3 categories: (d), (f) both have a flat shoulder in the low energy range; and (e), (g), (h) all have mainly two broad strong peaks; and (c) stands on its own because it has 3 major features. This 3-group classification is chemically intuitive : N3 and N5 (see panel (b) for labeling) belong to the \---NCS group; N4,N6,N7 are pyridine nitrogens and N2 is the only amine nitrogen atom. Without experimental support one has difficulty to judge which core hole model gives more reasonable DCH VtC-XES signals, but they differ qualitatively in spectral profile: in all curves calculated with the delocalized core hole model and the conventional SCH VtC-XES experiment, the higher energy peak is stronger than the lower energy peak (C$_0$ > B$_0$, C > B,...), while for the curves calculated with the localized core hole model, there are more cases of the lower energy peak stronger than the higher energy peak, thus the average spectrum has a stronger lower energy peak (see the dotted curve in panel(h), suppose all N atoms have equal chances for ionization). Checking the relative intensities of the major peaks in the higher and lower energy ranges in the experimental DCH XES spectra would give an easy test of both the delocalized and localized core hole models.
 
 \section{Conclusions and Brief Outlook}
In this paper we theoretically explored ts-DCH metal 1s/metal 1s  and metal 1s/ligand 1s DCH VtC-XES of representative transition metal model complexes. DCH VtC-XES is a new form of X-ray nonlinear spectroscopy enabled by the rapid development of XFELs. Our simulations show that through the perturbation introduced by a second core hole near the studied core hole, DCH VtC-XES can go beyond the conventional SCH VtC-XES techniques, and provide further information on the local electronic structure of the core holes and especially the interaction between the two atoms with core holes. In the near future, DCH VtC-XES has the potential to become a new research tool in transition metal complex chemistry and ultrafast science studies.

In this paper we have focused on only ts-DCH XES signals and not single-site (ss) DCH-XES, which could also be interesting. For deep ss-DCHs (\textit{e. g.}, 1s/1s ss-DCHs), the overall effect of the second core hole on the XES spectrum is mainly a constant shift of all emission lines and thus less interesting. However, 1s/2p ss-DCH-XES is more informative because the 2p core hole unlike the spherical 1s core hole, couples with the 3d and other valence orbital in different ways. In collaboration with Fuller et al., recently we observed Fe1s/Fe2p ss-DCH K$\alpha$ XES in Fe systems at the SACLA XFEL facility\cite{Fuller19}. In previous DCH photoelectron spectroscopy experiments\cite{Salen12}, XFEL pulses with a duration of $\sim$ 10 fs and an intensity over $10^{16}$ W/cm$^2$ were used. Since  hard X-ray core electron photoionization cross sections are generally one order of magnitude smaller than  soft X-ray core electron photoionization cross sections\cite{Scofield73}, we believe that even shorter and more intense XFEL pulses are needed for metal-metal DCH spectroscopy experiments.  Simulated experiments of metal/ligand DCH-XES in this study are still not currently available because of the difficulty of combining hard and soft XFEL pulses, but this technique could be possible with the planned Tender X-ray Instrument (TXI) in the under-constructing LCLS-II facility. Moreover, for metal/metal DCH-XES, the ultrashort lifetime of DCH states not only requires extremely short and intense XFEL pulses, but also broadens the emission lines. This line broadening issue could be remedied by using the recently developed stimulated XES technique\cite{Rohringer12,Yoneda15,Kroll18}, with which specific emission lines could be selectively enhanced and narrowed. Finally, VtC-XES is only the starting point of theoretical DCH spectroscopy. Fast and reliable relativistic quantum chemistry methods describing 2p core holes with spin-orbit coupling and  real-time simulations on ultrafast core hole dynamics are needed for a comprehensive understanding of other DCH spectroscopy techniques.

\begin{acknowledgments}

This material is based upon work supported by the U.S.
Department of Energy, Office of Science, Office of Basic Energy Sciences under the contract No. DE-AC02-76SF00515 (Y.Z, U.B., R.S.) including the Laboratory Directed Research and Development funding (Y.Z. and U.B.), and KC030105172685 (N.G.) and DE-SC0019277 (M.K.). This research was performed using EMSL, a DOE Office of Science User Facility sponsored by the Office of Biological and Environmental Research and located at PNNL. PNNL is operated by Battelle Memorial Institute for the United States Department of Energy under DOE contract number DE-AC05-76RL1830. The research also benefited from resources provided by the National Energy Research Scientific Computing Center (NERSC), a DOE Office of Science User Facility supported by the Office of Science of the U.S. Department of Energy under Contract No. DE-AC02-05CH11231.

\end{acknowledgments}

\providecommand{\latin}[1]{#1}
\providecommand*\mcitethebibliography{\thebibliography}
\csname @ifundefined\endcsname{endmcitethebibliography}
{\let\endmcitethebibliography\endthebibliography}{}

\end{document}